\documentclass{elsart}
\usepackage{natbib}
\usepackage{graphics}
\usepackage{amssymb}

\begin{document}

\begin{frontmatter}
\title{SPACE STORM MEASUREMENTS OF THE JULY 2005 SOLAR EXTREME EVENTS FROM THE LOW CORONA TO THE EARTH}
\author[DI]{C. Caroubalos},
\author[UOA]{P. Preka-Papadema},
\author[NPH]{H. Mavromichalaki},
\author{X Moussas\corauthref{UOA}}
\corauth[UOA]{X Moussas} \ead{xmoussas@phys.uoa.gr},
\author[NPH]{A. Papaioannou},
\author[UOA]{E. Mitsakou}
\and
\author [UOA]{A. Hillaris}

\address[DI]{Department of Informatics, University of Athens, GR-15783 Athens, Greece}
\address[UOA]{Section of Astrophysics, Astronomy and Mechanics, Department of Physics, University of
Athens, GR-15784 , Panepistimiopolis Zografos, Athens, Greece}
\address[NPH]{Section of Nuclear and Particle Physics, Department of Physics, University of
Athens, GR-15784 , Panepistimiopolis Zografos, Athens, Greece}

\begin{abstract}
The Athens Neutron Monitor Data Processing (ANMODAP) Center recorded an unusual Forbush decrease with a sharp enhancement of cosmic ray intensity right after the main phase of the Forbush decrease on 16 July 2005, followed by a second decrease within less than 12 hours. This exceptional event is neither a ground level enhancement nor a geomagnetic effect in cosmic rays. It rather appears as the effect of a special structure of interplanetary disturbances originating from a group of coronal mass ejections (CMEs) in the 13-14 July 2005 period. The initiation of the CMEs was accompanied by type IV radio bursts and intense solar flares (SFs) on the west solar limb (AR 786); this group of energetic phenomena appears under the
label of {\em{Solar Extreme Events of July 2005}}. We study the characteristics of these events using combined data from earth (the ARTEMIS IV radioheliograph, the Athens Neutron Monitor (ANMODAP)), space (WIND/WAVES) and data archives. We propose an interpretation of the unusual Forbush profile in terms of
a magnetic structure and a succession of interplanetary shocks interacting with the magnetosphere.
\end{abstract}

\begin{keyword}
Coronal Mass Ejections \sep
Flares \sep
Activity \sep
X-Rays \sep
Cosmic Rays \sep
Forbush decreases


\end{keyword}
\end{frontmatter}
\section{INTRODUCTION}\label{Intro}
Space weather drivers, such as CMEs, energetic particles and MHD shocks
are mostly of solar origin; these modulate the flux of galactic cosmic rays in the form of Forbush decreases.

Solar radio bursts, on the other hand, provide an extremely efficient diagnostic of the drivers
onset in the corona. The type II bursts are MHD shocks;
a subset of them are driven by CME and manifest the coronal origin of interplanetary shocks.
The type IV continua originate from energetic electrons trapped within plasmoids, magnetic
structures or substructures of CMEs (\citeauthor{Bastian} \citeyear{Bastian}); those often indicate mass ejection
and propagation in the low corona. Lastly, the type III bursts trace the propagation of energetic
electrons through the corona and, often, mark the onset of energy release processes.
{\begin{figure*}[ht]
\resizebox{\hsize}{!}{\includegraphics{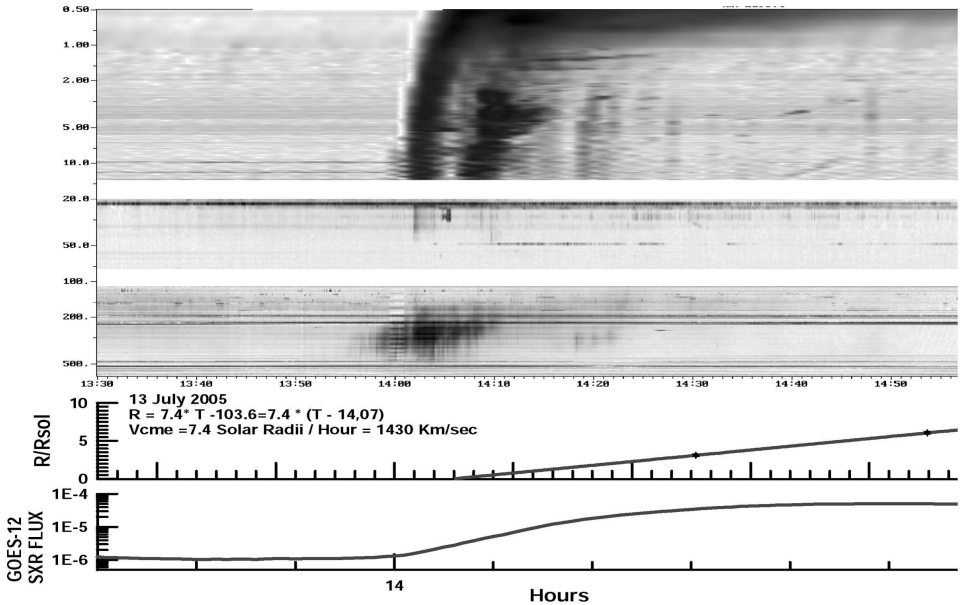}}
\caption{Top: Composite ARTEMIS--IV/WIND Dynamic Spectrum of the 13 July 2005 Event: A type IV
continuum, 14:01--14:30 UT, and a group of type III electron beams drifting from the
low corona to the WIND/WAVES range.
Middle: Height Time plot of the CME; the least squares fit indicates CME take-off at 14:04 UT
at plane of the sky speed 1440Km/sec. Bottom: The GOES SXR light curve}
\label{DynSpec13}
\end{figure*}}
{\begin{figure*}[ht]
\resizebox{\hsize}{!}{\includegraphics{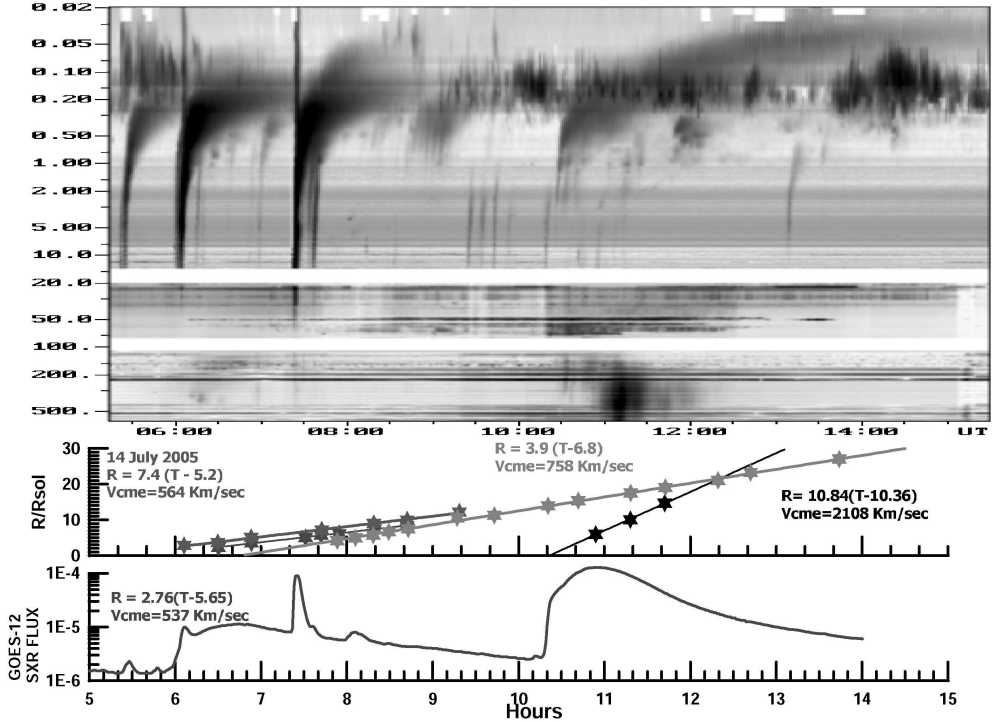}}
\caption{Top: Composite ARTEMIS--IV/WIND Dynamic Spectrum of the 14 July 2005 Event:
Type III groups extending into the WAVES range;
a faint  continuum, 06:00--06:30 UT, and a type IV burst, 10:20--12:20 UT.
Middle: Height Time plot of the fast CME overtaking the slow ones.
Bottom: The GOES SXR light curve}
\label{DynSpec14}
\end{figure*}}
{\begin{figure*}[ht]
\resizebox{\hsize}{!}{\includegraphics{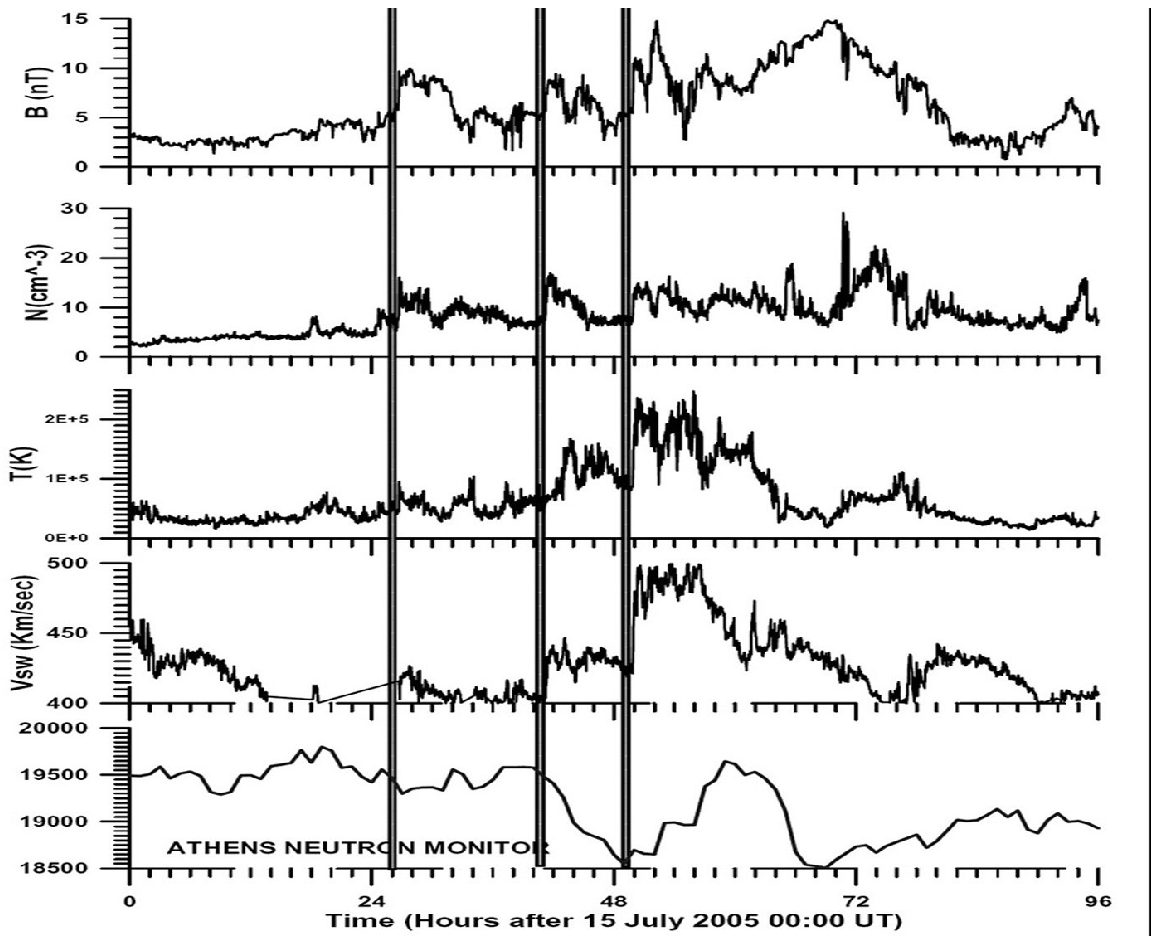}}
\caption{Solar wind parameters from OMNI; Top to Bottom: Total magnetic field strength,
proton density, proton temperature and Solar Wind Speed. Three weak shocks are marked
(cf. discussion in text)
Bottom Panel: The ANMODAP plot of the irregular Forbush decrease, interrupted by a
sudden enhancement.}
\label{NMonitor}
\end{figure*}}

A group of energetic solar phenomena were observed on the Sun in active region 786 (N$10^o$ W$90^o$)
on 13-14 July 2005.  The associated magnetospheric response affected cosmic ray
(Neutron Monitor) measurements and space weather from 16 to 18 July, marking this activity as the
{\em extreme events of July 2005}. On the 16 July, in particular, a sharp decrease in cosmic ray count
rate was recorded yet right after the minimum an enhancement followed, distorting the typical
profile of a Forbush decrease. The peculiarity of this event and the conditions under which it originated are examined.

\section{OBSERVATIONAL DATA \& ANALYSIS} \label{Obs}
\subsection{Data sets}\label{Data}
The data sourses used in our analysis were:
\begin{itemize}
\item{The Artemis--IV\footnote{Appareil de Routine pour le Traitement et l' Enregistrement
Magnetique de l' Information Spectral} (\cite{Caroubalos} also \cite{Kontogeorgos, Kontogeorgos08}) 
radiospectrograph at Thermopylae (http://www.cc.uoa.gr/artemis/);
it covers the frequency range from 650 to 20 MHz with time resolution of 0.1 s.}
\item{ The WIND/WAVES receivers (\citeauthor{Bougeret}, \citeyear{Bougeret}) 
in the range 20kHz--14 MHz, complement the
spectral range of ARTEMIS--IV; the combined observations are used in the study of the
continuation of solar bursts in the interplanetary space bridging thus the gap between space borne
and ground-based radio observations.}
\item{CME data from the LASCO lists on line (http://cdaw.gsfc.nasa.gov/CME{$\_$}list, \cite{Yashiro})}
\item{The Nan\c cay Radio heliograph (\citeauthor{Kerdraon}, \citeyear{Kerdraon}) 
for positional information of radio emission.}
\item{ SXR (GOES) on line records (http//www.sel.noaa.gov/ftpmenu/indices).}
\item{The Neutron Monitor Station of Athens University (\citeauthor{Mav01}, \citeyear{Mav01})
and the corresponding data analysis center (http://cosray.phys.uoa.gr, \cite{Mavromichalaki})}
\item{Solar wind parameters from the OMNI (http://omniweb.gsfc.nasa.gov/) online database.}
\end{itemize}
\subsection{Solar Activity Observations}\label{Solar}

All solar energetic phenomena studied originated in AR 786 (N$10^o$ W$90^o$).

The Solar activity on the 13 July 2005 starts at 14:01 UT with an M5.0 long duration SXR flare
ending at 15:38 UT. A fast halo CME, with speed 1430 km/sec
takes off at 14:12 UT. From the ARTEMIS/WIND recordings we establish that a type
IV burst (14:01-14:30 UT) overlaps in time with the flare onset and the estimated
CME lift--off (cf. figure\ref{DynSpec13});
from the Nan{\c c}ay Radio heliograph images the position of the continuum appears over AR 786.

The active phenomena of the 14 July 2005 commence with an M9.1 (05:57--07:43 UT)
flare followed by an X1.2 (10:16-11:29 UT).
The former is associated with type III groups and the lift--off
of three slow CMEs, with estimated take off at 05:32 UT, 06:01 UT and 07:02 UT and corresponding
speeds 514, 573 \& 758 Km/sec; a sharp SXR peak associated with a group of type III bursts and
an SF H${\alpha}$ flare (reported by LEARMONTH) appears at about 07:59--08:12 UT
although is not included in the GOES SXR flare lists.

All CMEs start with almost the same position angle ($252^o$--$282^o$)
and with increasing width ($14^o$, $60^o$, $103^o$); those appear as successive ejections from
AR 786 which eventually merge as the faster overtake the slower. This activity is accompanied by a
faint continuum from 06:00 UT to 06:30 UT in the 500--100 MHz range.
The X1.2 flare is associated with a fast CME (2108 Km/sec) which takes off at 10:27 UT overtaking the
three slow CMEs at about 12:20 UT. It is also accompanied by type III bursts and
a type IV continuum (10:20--12:20 UT).

In figure \ref{DynSpec14} an overview of the active phenomena of the 14 July 2005 is presented.

\subsection{Solar Wind Parameters Analysis - Effects on Cosmic Ray Modulation}\label{CR}
A large Forbush decrease (9\% - at south polar stations) and sharp changes of the anisotropy
occurred on 16--17 July 2005; these more or less coincide with medium level disturbances in the interplanetary space (Figure \ref{NMonitor}) which, in turn, correspond to weak interplanetary shocks
without coronal counterparts.

The shocks were
recorded in the OMNI data base (July 16, 02:35 UT, 17:06 \& July 17 01:41 UT) and appeared at the
near--Earth space 2--2.5 days after the fast CME onsets of the
13--14 July, therefore they are expected to be driven by them; we note that their time difference is about 23 hours while the interval between successive fast CMEs was about 20.
The passage of each interplanetary shock was marked by an increase in magnetic field strength
(5.8 to 8.0 nT, 5.2 to 8.0 nT and 5.6 to 9.3 respectively), an increase in proton density (6.7 to 11.10 cm$^{-3}$, 6.7 to 13.7 cm$^{-3}$ \& 6.7 to 11.10 cm$^{-3}$) and temperature (47800 K to 63100K and
subsequently to 165300 K). The variation in the solar wind speed shows rather small changes (Figure
\ref{NMonitor}), implying that only a small part of the mass ejection interacted with the Earth's
magnetosphere as the CME was launched from the limb.
The shock speeds were computed at the Earth's orbit from v:=(n$_2$v$_2$-n$_1$v$_1$)/(n$_2$-n$_1$),
where n$_1$, v$_1$ and n$_2$, v$_2$ the upstream and downstream plasma density and velocity respectively
and v the shock velocity. The calculated speeds were found to be 509, 434 and 557 Km/sec
exceeding the solar wind speed values reported in OMNI data base which were 420, 411, 483 Km/sec respectively.

The direction of B as reported in the OMNI data base is found to be south (Bz$<$0) for the first and in part the third IP shock; this is consistent with a small variation of the geomagnetic field (Kp index)
and a double sub storm of -60 and -76 nT (Dst index) which were also recorded in the same data set.
This event cannot be classified as strong;
were this the case the Dst index should be lower than -100 nT (\citeauthor{Loewe}, \citeyear{Loewe}) resulting
in a strong geomagnetic storm according to the NOAA Space Weather
Scales (http://www.swpc.noaa.gov/NOAAscales/).

An intensive Forbush decrease of cosmic rays, recorded on the 16th of July, was observed by the majority of the neutron monitors worldwide.
After the main phase of the Forbush decrease at the 17th of July, a sharp enhancement of cosmic ray
intensity occurred and was followed by a second decrease, within 8 hours (cf. Figure \ref{CosmicRays}).
An unusually high anisotropy of cosmic rays ($\approx$ 7-8 \%) was observed,
especially of the equatorial component with a direction to the western source. The
north-south anisotropy of cosmic rays was also extensively large, $\approx$ 7\%,
as reported by \cite{Mavromichalaki1} and \cite{Papaioannou}.
{\begin{figure*}[ht]
\resizebox{\hsize}{!}{\includegraphics{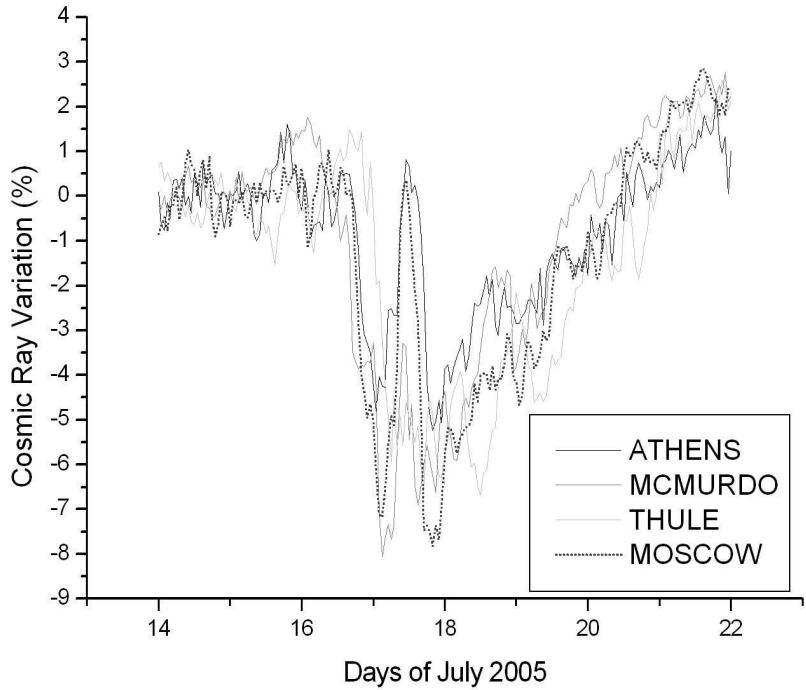}}
\caption{Cosmic ray variations from 14 to 22 July 2005 from various stations: Athens, McMurdo, Thule and Moscow}
\label{CosmicRays}
\end{figure*}}

The peculiarity of this event is due to the fact that it is neither a ground level enhancement of solar cosmic rays nor a geomagnetic effect in galactic cosmic rays. It rather appears as
the result of solar modulation of galactic cosmic rays, which were recorded under relatively quiet geomagnetic conditions; this also precludes the {\em{rogue event}} case (\citeauthor{Kallenrode}, \citeyear{Kallenrode}).

It seems that the sequence of the CMEs and the corresponding interplanetary shocks have produced the above cosmic ray decrease with the two distinct deep minima;
the time between them is approximately 20 hours, as it is the time interval between successive interplanetary shocks, and between the fast CME lift--of on the 13th \& 14th of July respectively.

\section{DISCUSSION AND CONCLUSIONS}\label{AR}
In this report the solar extreme events on the 13 and 14 July 2005 and the
associated Forbush decrease has been studied. The initiation of CMEs is linked to the 
appearance  of type IV radio bursts and strong
solar flares. Their effects were traced from the base of  the solar corona to the Earth; 
they included complex radio bursts and variations in cosmic ray fluxes and space weather.

Three interplanetary shocks were observed about 48 hours after the CMEs lift off; their time intervals 
in both cases were similar (about 20 hours) yet they were not accompanied by significant change of the 
solar wind parameters, the solar wind speed in particular, probably due to the origin of the CMEs on the west limb. A substorm (Dst double minima of -60 and -76 nT) and a Forbush decrease (double minima with a time interval of 20 hours between them) were recorded; 
this time interval, between minima, was consistent with the time between the IP shocks.

The peculiar cosmic ray behavior was registered mainly at mid-latitude (Athens, Moscow) and south polar (McMurdo) neutron monitor stations; at north polar stations (Thule) the forbush decrease
(amplitude  $\approx$ 7\%) did not exhibit the peculiar 
double minimum under study (cf. Figure \ref{CosmicRays}). This is probably connected to the high north-south anisotropy of cosmic rays mentioned in subsection \ref{CR}.

It appears that Earth was influenced by the first shock which initiated the
Forbush decrease on the 16th-17th of July. After the minimum, the Earth
moved outside the shock magnetic structure and thus galactic cosmic rays were, once more,
recorded. On the same day, 17th of July, the third shock initiated the second part of this irregular Forbush decrease. This may account for the double minimum an the said irregularity.

A similar proposition, regarding the event of October 28, 2003, appears in \cite{Miros} where
special interplanetary conditions may affect cosmic rays.

\ack{This work was financially supported by the Special Research Committee of the University of Athens and by PYTHAGORAS II project which is funded by European Social Funds and National Resources. The authors thank the referees for suggestions which improved the quality of the paper.}

\end{document}